\documentclass[twocolumn]{aastex62}

\usepackage{mathtools}
\usepackage{amsmath}
\usepackage{subfigure}
\usepackage{bm}
\usepackage{ulem}
\newcommand{\be}{\begin{equation}}
\newcommand{\ee}{\end{equation}}
\newcommand{\ba}{\begin{eqnarray}}
\newcommand{\ea}{\end{eqnarray}}
\newcommand{\bi}{\begin{itemize}}
\newcommand{\ei}{\end{itemize}}
\newcommand{\no}{\nonumber}

\usepackage{color}

\usepackage{ulem}

\usepackage{graphicx} 
\begin{document}

\title{The copula of the cosmological matter density field is non-Gaussian}

\author{Jian Qin}
\affiliation{Department of Astronomy, School of Physics and Astronomy, Shanghai Jiao Tong University, Shanghai, 200240,China}
\affiliation{Shanghai Key Laboratory for Particle Physics and Cosmology, Shanghai, 200240, China} 

\author[0000-0002-9359-7170]{Yu Yu}
\affiliation{Department of Astronomy, School of Physics and Astronomy, Shanghai Jiao Tong University, Shanghai, 200240,China}
\affiliation{Shanghai Key Laboratory for Particle Physics and Cosmology, Shanghai, 200240, China}  

\author{Pengjie Zhang}
\affiliation{Department of Astronomy, School of Physics and Astronomy, Shanghai Jiao Tong University, Shanghai, 200240,China} 
\affiliation{Division of Astronomy and Astrophysics, Tsung-Dao Lee Institute, Shanghai Jiao Tong University, Shanghai, 200240, China}
\affiliation{Shanghai Key Laboratory for Particle Physics and Cosmology, Shanghai, 200240, China} 

\email{yuyu22@sjtu.edu.cn, zhangpj@sjtu.edu.cn}

\begin{abstract}
Non-Gaussianity of the cosmological matter density field can be largely reduced by a local Gaussianization transformation (and its approximations such as the logarithmic transformation). Such behavior can be recasted as the Gaussian copula hypothesis, and has been verified to very high accuracy at two-point level.  On the other hand,  statistically significant non-Gaussianities in the Gaussianized field have been detected in simulations.  We  point out that, this apparent inconsistency is caused by the very limited degrees of freedom in the copula function, which make it misleading as a diagnosis of residual non-Gaussianity in the Gaussianized field.  Using the copula density {\color{black}{{and at the two-point level}}}, we highlight the departure from Gaussianity. We further quantify its impact in the predicted $n$-th ($n\ge 2$) order correlation functions.   We explore a remedy of the Gaussian copula hypothesis, which alleviates but not completely solves the above problems. 
\end{abstract}


\section{Introduction} \label{sec:intro}
The large scale structure (LSS) fields in the late-time universe, in particular the matter density field,  are often significantly non-Gaussian, due to the nonlinear evolution of the universe \citep{2002PhR...367....1B}.  It is an active research frontier to describe the non-Gaussianity accurately and to extract the encoded cosmological information efficiently \cite[e.g.,][]{2000MNRAS.312..257H,2011ApJ...728...35Z,2012MNRAS.421..832Y}. An interesting finding is that, despite the vast possibility of non-Gaussian behaviors, the non-Gaussianity of the matter density field induced by nonlinearity takes a specific form of simplicity. It has been known for a long time that a local monotonic transformation of the density field ($\delta({\bf x})\rightarrow y({\bf x})=f(\delta({\bf x}))$) can significantly reduce the non-Gaussianity. Namely,  the non-Gaussianity is largely encoded in the one point probability distribution function (PDF), and the field ($y({\bf x})$) after the above local transformation is close to Gaussian.  In the literature, various approximations such as the logarithmic transform \citep{1991MNRAS.248....1C,Neyrinck2009}, the rank-order transform  \citep{1992MNRAS.254..315W,Neyrinck2011,2016MNRAS.457.3652M},   the Box-Cox transform \citep{2011MNRAS.418..145J},  and the clipping method \citep{2011PhRvL.107A1301S}, along with the exact Gaussianization transformation \citep{2011PhRvD..84b3523Y}, have been investigated. All are able to significantly reduce the non-Gaussianity and enhance the information content encoded  in the two-point statistics.  

Copula provides an alternative description on the above findings.  The $n$-point PDF $f(\delta_1,\delta_2, \dots, \delta_n)$ ($n=1,2,\dots$) completely describes  the statistics of the  density field. They can be equivalently described by the combination of one-point PDF $f(\delta)$, and all the $n$-point copula $C(u_1, u_2, \dots, u_n)$ ($n\geq 2$).  Here $u(\delta)\equiv F(\delta)\equiv \int_{-\infty}^\delta f(\delta^{'})d\delta^{'}$, and $F(\delta)$ is the cumulative distribution function of the density field.  Copula has a nice property, that it is invariant under any local monotonic transformation. This makes it convenient to describe the residual non-Gaussianity in the $y$ field. For example, if  the copula is found to depart from the Gaussian form, then no local transformation can render the density field Gaussian. 
\cite{Scherrer2009} found through N-body simulations that the two-point copula for all the investigated spatial separation is Gaussian to extremely high accuracy. This motivates the authors to  
postulate the Gaussian copula hypothesis (GCH), that all $n$-point copulas are Gaussian. 
The GCH, along with the one-point (non-Gaussian) PDF, provides a convenient and close form description of the non-Gaussian density field. It has been applied to study the covariance matrix of lensing power spectrum and other statistics \cite[e.g.,][]{2010PhRvL.105y1301S,2010MNRAS.406.1830T,2011PhRvD..83b3501S,2015A&A...583A..70L,2016PhRvD..94h3520Y, 2018ApJ...869...74Z}.

However, direct investigation of the Gaussianized field (the $y$ field) shows the existence of residual non-Gaussianity. For example, non-vanishing bispectra \citep{2011PhRvD..84b3523Y}) and off-diagonal covariance matrix elements of the power spectrum \citep{2016PhRvD..94h3520Y} have been detected robustly. 
An intuitive illustration of the residual non-Gaussianity is the strong anisotropic structures in the two-dimensional visualizations of the Gaussianized density fields (e.g., Figure 1 in \cite{Neyrinck2009}). 

What causes the above inconsistency? One possibility is the non-Gaussianity in higher order copula (e.g. $n\geq 3$). Here we point out an alternative possibility.  Two-point copula investigated in \cite{Scherrer2009} is indeed nearly Gaussian, as we have verified independently with our simulations. However this nearly Gaussian copula is misleading due to some build-in nature of copula. We take the two-point copula $C(u,v)$ as an example. It is subject to the following constraints,
\ba
\label{eqn:constraints}
\begin{aligned}
&u,v\in [0,1] \ ; C(u,v)\in[0,1]\  ; C(u,v)=C(v,u);\ \\
&\frac{\partial C(u,v)}{\partial u}\geq 0\ ;\ \frac{\partial C(u,v)}{\partial v}\geq 0 \ ; \\
&C(u,0)=C(0,v)=0\ ; \ C(u,1)=u\ ;\ C(1,v)=v\ . 
\end{aligned}
\ea
So it has very limited degrees of freedom. It monotonically increases with both $u$ and $v$. It has fixed values when $u=0,1$ or $v=0,1$. So different copulas may look similar. This has two implications.  (i) For example, from the viewpoint of LSS, a field with significant spatial correlation, and a random field of vanishing spatial correlation, are fundamentally different. However, the two copulas are almost identical in the vicinity of $(u,1)$, $(u,0)$, $(0,v)$ and $(1,v)$. This implies that even if the copula is close to Gaussian, the field may still have significant non-Gaussianity.  (ii) Also for the same reason,  tiny difference in copula may result into significant difference in the more commonly used correlation function and high order correlations.  These statistics are not investigated in \cite{Scherrer2009}.  However, as derived properties from the copula and one-point PDF, they can serve to quantify the accuracy of Gaussian copula hypothesis.

For further check of  the first implication, we adopt the copula density,  the partial derivatives of the copula. It is no longer subject to the constraints that the copula is subject to Equation (\ref{eqn:constraints}). We find that,{\color{black}{{ at the two-point level, the copula density}}} reveals clearly some non-Gaussianities otherwise hidden deeply in the copula. For the second, we explicitly calculate and compare $\xi_{mn}\equiv \langle \delta_1^m\delta_2^n\rangle-\langle \delta_1^m\rangle\langle\delta_2^n\rangle$ in N-body simulations. We find that GCH fails, even for the lowest order case ($m=n=1$).  Given the convenience of GCH in calculating LSS statistics, we explore a remedy of GCH.  Two-point Gaussian copula has a single free parameter $r$ and GCH fixes it to a specific value. We have the freedom to adopt different values, while keeping the Gaussian form\footnote{If a copula of a random field is Gaussian, the parameters in the Gaussian copula could be uniquely determined by the properties of Gaussian distribution \citep{2003QuFin...3..231M}. If not, Gaussian copula with parameters determined appropriately could be regarded as approximation to the real non-Gaussian copula for research interest.}. This alternative Gaussian copula approximation improves the description of $\xi_{mn}$, but fails at $10\%$ accuracy level for $m+n\geq 2$. This further demonstrates the failure of GCH. 

This paper is organized as follows. In Section 2 we briefly introduce the copula. Section 3 describes the steps to reveal  non-Gaussianity of the {\color{black}{{two-point}}} copula density. We show the significant bias in correlation functions by almost invisible deviation from Gaussian copula.  In Section 4 we investigate an alternative Gaussian copula approximation to improve the correlation function statistics. In Section 5 we discuss and summarize the results.

\section{Preliminaries of the Copula Function } \label{sec:copula}
For overdensity $\delta_i\equiv \delta({\bf x}_i)$ ($i=1,2,\dots,n$), the copula function $C(u_1,u_2,\dots, u_n)$ is defined as
\begin{equation}
\label{copdef}
F(\delta_1, \delta_2,\dots,\delta_n) = C(F(\delta_1),F(\delta_2),\dots,F(\delta_n))\ .\\
\end{equation}
Here, $u_i\equiv F(\delta_i)$, and  $F(\delta_i)$ is the marginal cumulative distribution function (CDF) of $\delta_i$. Namely $u_i\equiv \int_{-1}^{\delta_i} f(\delta)d\delta$ and $f(\delta)$ is the probability distribution function of the density field.  $F(\delta_1, \delta_2,\dots,\delta_n)$ is the joint cumulative distribution function (JCDF) of $\delta_1,\delta_2,\dots,\delta_n$. 
One important property is that for any random field, the copula function defined by Equation (\ref{copdef}) always exists and it is unique \citep{Sklar1}.

If $C$ is differentiable, then 
\begin{equation}
\begin{aligned}
 f(\delta_1,\delta_2,\dots,\delta_n)
 &\equiv \frac{\partial^n F(\delta_1,\delta_2,\dots,\delta_n)}{\partial\delta_1\partial\delta_2\dots\partial\delta_n}\\
 &=\frac{\partial ^n C(u_1, u_2, \dots,u_n)}{\partial u_1\partial u_2 \dots \partial u_n}{\displaystyle\prod_{i=1}^{n} \frac{\partial F(\delta_i)}{\partial\delta_i}}\\
&=c(u_1, u_2, \dots, u_n){\displaystyle\prod_{i=1}^{n} {f(\delta_i)}}\ .\\
\label{pdfgcd}
\end{aligned}
\end{equation}
The copula density is defined as 
\begin{equation}
\begin{aligned}
 c(u_1, u_2, \dots, u_n)\equiv \frac{\partial ^n C(u_1, u_2, \dots,
u_n)}{\partial u_1\partial u_2 \dots \partial u_n} \ .
\end{aligned}
\end{equation}
It is related to the joint PDF (JPDF) $f(\delta_1, \delta_2, \dots,\delta_n)$ by 
\begin{equation}
\begin{aligned}
 c(u_1, u_2, \dots, u_n)=\frac{f(\delta_1,\delta_2,\dots,\delta_n)}{\displaystyle\prod_{i=1}^{n} {f(\delta_i)}}\ .
\label{cdjpdf}
\end{aligned}
\end{equation}

\begin{figure*}[ht!] 
\vspace{-2cm}
\centering
\includegraphics[width=19cm]{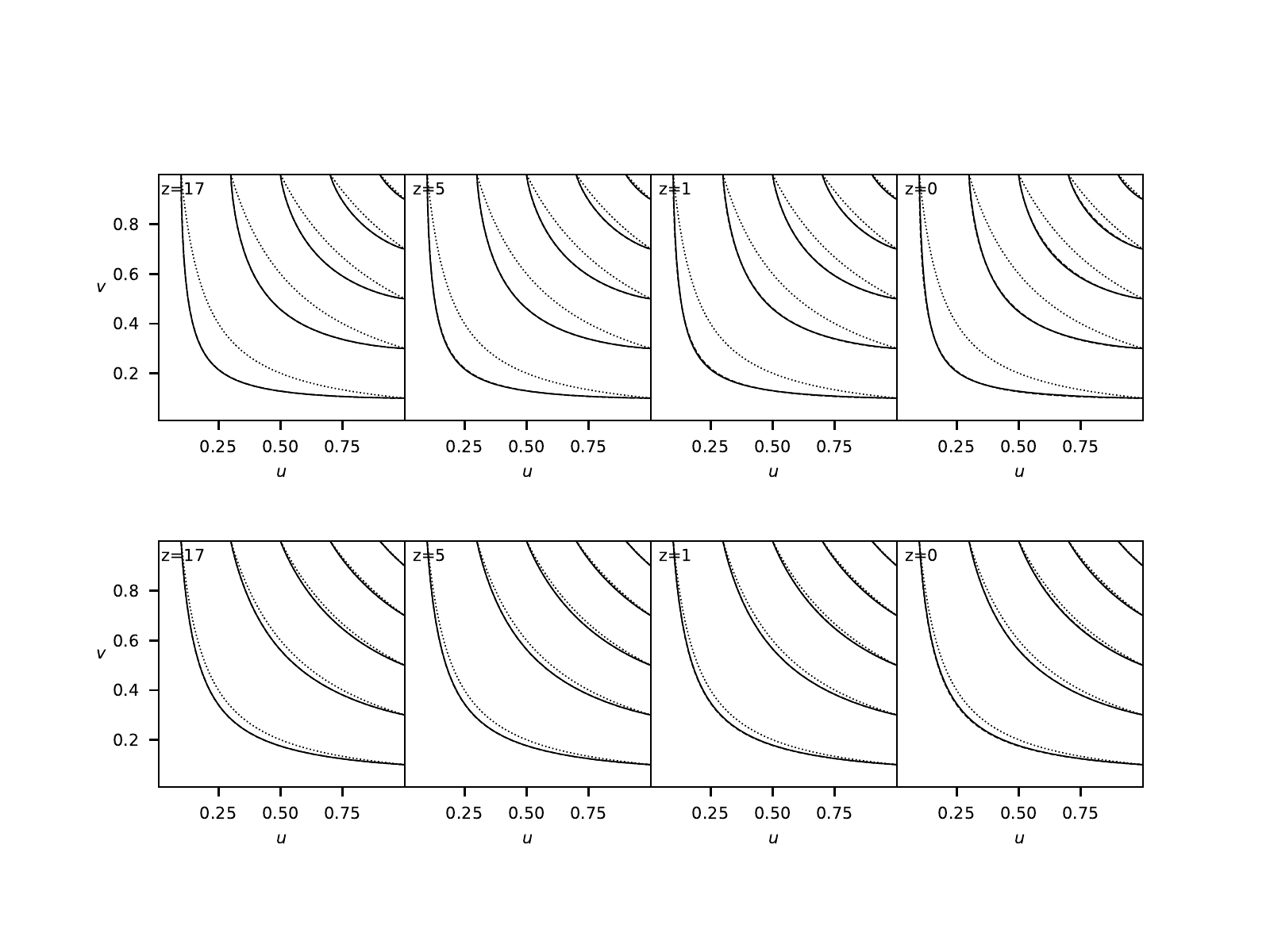}
\vspace{-1.5cm}
\caption{Empirical two-point copula $C_D(u,v)$ for the simulated dark matter density distributions at the two-point separations 2Mpc$/h$ (top panels) and 6 Mpc$/h$ (bottom panels). The four columns show results at redshifts $z=17,5,1,0$. Solid curves are the contours corresponding to (from lower left to upper right) $C_D(u,v)=0.1, 0.3, 0.5, 0.7, 0.9$. The dashed curves show the Gaussian copula measured based on GCH. { Notice that they are almost indistinguishable from the solid curves.  The dotted curves show the copula of spatially uncorrelated fields. Notice the similarity in the copula, especially in the vicinity of $u\rightarrow 0,1$ or $v\rightarrow 0,1$. Such similarity implies limitations of copula in describing LSS, and in particular the LSS non-Gaussianity. }
\label{Cuv}}
\end{figure*}

\begin{figure*}[ht!]
\vspace{-2cm}
\centering
\includegraphics[width=20cm]{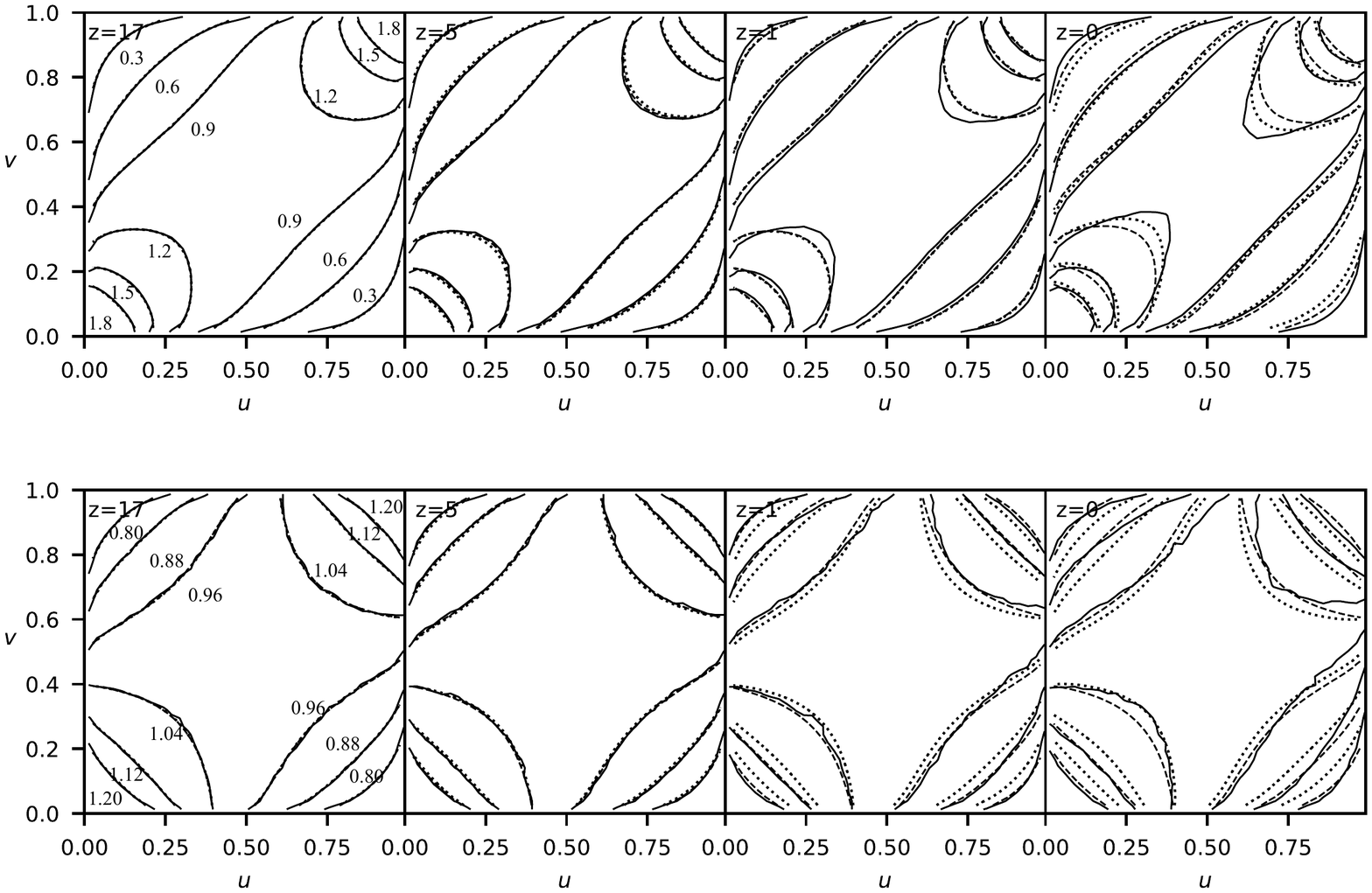}
\vspace{-2cm}
\caption{Empirical two-point copula densities $c_D(u,v)$ for the simulated dark matter density distributions at the two-point separations 2Mpc$/h$ (top panels) and 6 Mpc$/h$ (bottom panels). The four columns show results at redshifts $z=17,5,1,0$. Solid curves are the contours corresponding to $c_D(u,v)=1.8,1.5,1.2,0.9,0.6,0.3$ (top panels) and $c_D(u,v)=1.20,1.12,1.04,0.96,0.88,0.80$ (bottom panels). The dashed curves give the Gaussian copula densities measured based on GCH, and the dotted curves give the results of the alternative Gaussian copula approximation.
}
\label{cuv}
\end{figure*}

\begin{figure*}[ht!]    
\centering
\includegraphics[width=19cm]{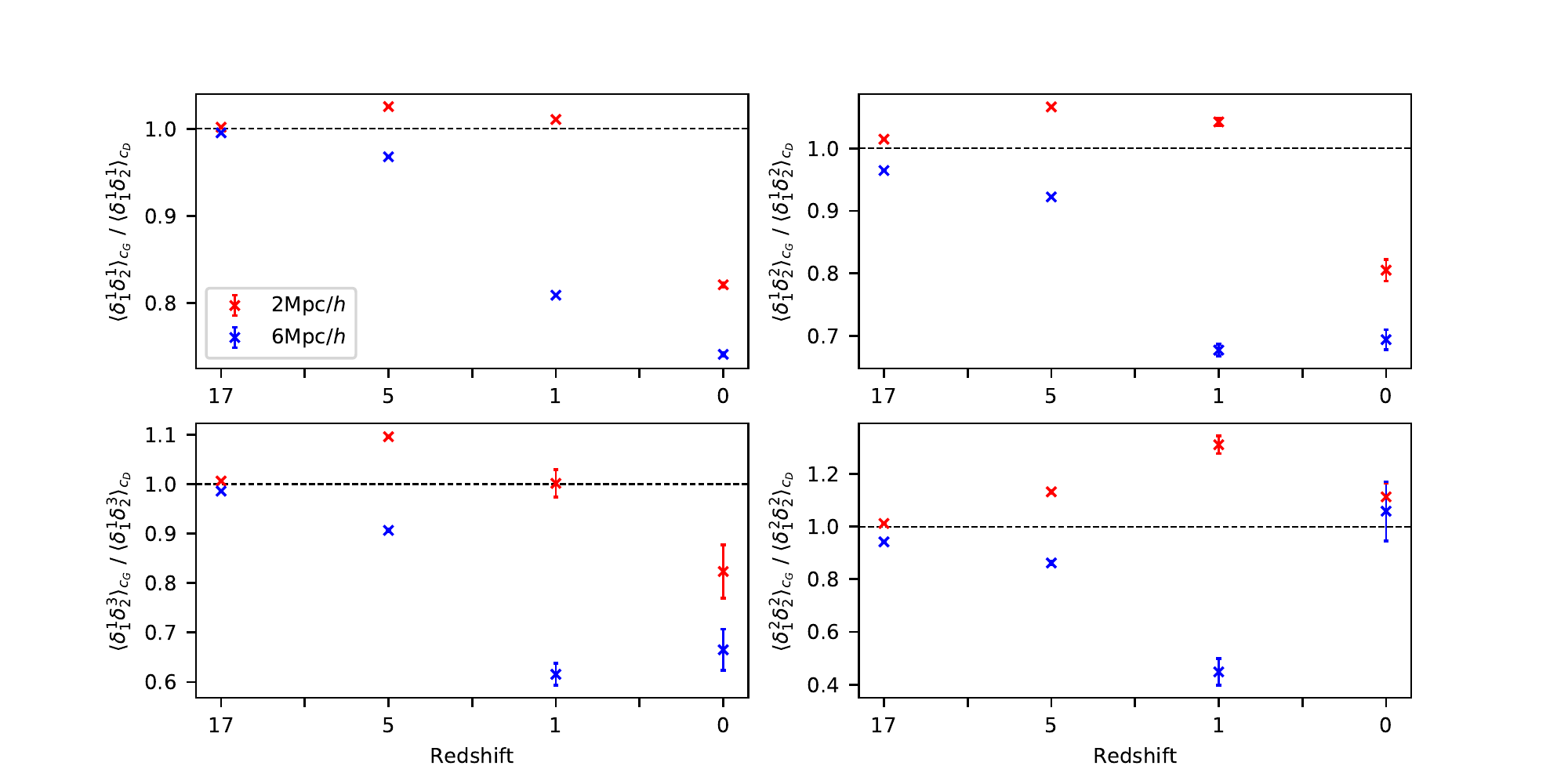}   
\caption{
Comparison of correlation functions $\langle\delta_1^m\delta_2^n\rangle$ as a function of redshift, measured from the Gaussian copulas $\langle\delta_1^m\delta_2^n\rangle_{c_G}$ and from the simulated density fields $\langle\delta_1^m\delta_2^n\rangle_{c_D}$. 
From the top left panel to the bottom right panel, four cases (from $m=n=1$ to $m=n=2$) have been shown.  
Red points show the ratio of $\langle\delta_1^m\delta_2^n\rangle_{c_G}$ to $\langle\delta_1^m\delta_2^n\rangle_{c_D}$ for separation $s=2$ Mpc$/h$; blue points show the ratio for separation $s=6$ Mpc$/h$, with errorbars indicating the shot noise.
The Gaussian copulas are determined based on the GCH. 
}
\label{GCHbias1}
\end{figure*} 

\subsection{General Properties}
By the above definitions and results, we can derive some basic properties of copula. For brevity, we demonstrate them with  the 2-point copula $C(u,v)$.  
\bi
\item $u,v\in [0,1]$, and $C\in [0,1]$.
\item $C(u,v)=C(v,u)$. Notice that this holds for statistically homogenous fields such as the cosmological matter density field. It does not hold for general fields. 
\item $C(u,0)=0$ and  $C(u,1)=u$, as shown by Equation (\ref{copdef}). For the same reason, $C(0,v)=0$ and  $C(1,v)=v$. These properties hold for any fields, statistically homogenous or not. 
\item $\partial C/\partial u\geq 0$ and $\partial C/\partial v\geq 0$. Namely $C$ monotonically increases with both $u$ and $v$. This can be derived from Equation (\ref{cdjpdf}), which leads to $c\geq 0$. Therefore for any random fields, $C(u,v)$ always increases from $0$ at $v=0$ to $u$ at $v=1$, for fixed $u$. Due to this constraint,  different random fields can have similar $C(u,v)$.
\item Invariance of copula under monotonically increasing transformation $y=f(\delta)$. This is obvious since $F_y(y_1, y_2, \dots,y_n)=F_\delta(\delta_1, \delta_2, \dots,\delta_n)$. 
\ei

For special cases, the copula function has analytical expression. One is the case of two uncorrelated variables $\delta_1,\delta_2$. Since $F(\delta_1,\delta_2)=F(\delta_1)F(\delta_2)$, we have
\begin{equation}
C(u,v)=uv,\ \ c(u,v)=1\ .
\label{c=1}
\end{equation}
Another case is the Gaussian copula for the Gaussian field, as detailed below. 

\subsection{Gaussian Copula}
To distinguish the Gaussian CDF from a general CDF, we denote it as $\Phi_{\bm{\rho},n} \left(\delta_1,\dots, \delta_n \right)=F_G\left(\delta_1,\dots, \delta_n \right)$. The covariance matrix
\ba
\bm{\rho}_{ij}\equiv \langle \delta_i\delta_j\rangle\ ,\ {\rm with}\ \sigma_i^2\equiv \langle \delta_i^2\rangle\ ,
\ea
completely fixes the statistics of the Gaussian field. From Equation (\ref{copdef}), we derive the Gaussian copula \citep[see e.g.,][]{2003QuFin...3..231M,Neyrinck2011}
\begin{equation}
\label{eq:defgc}
C_{\rm G}(u_1, \dots, u_n) = \Phi_{\bm{\rho},n} \left(\Phi_1^{-1}(u_1),
\dots, \Phi_1^{-1}(u_n) \right)\ .
\end{equation}
Here $\Phi_1$ is the marginal Gaussian CDF,
\begin{equation}
 \Phi_1(x_i)=\int_{-\infty}^{x_i}\frac{1}{\sqrt{2\pi}\sigma_i}\exp\left(-\frac{x^2}{2\sigma_i^2}\right)d x\ .
\end{equation}
The Gaussian copula density is
\ba
\label{eq:gcd}
\begin{aligned}
c_{\rm G}(u_1, u_2, \dots, u_n)&=\frac{\sigma_1\sigma_2\cdots\sigma_n}{\sqrt{\det {\bf \bm{\rho}}}} \\ 
&\times \exp \left( -\frac{1}{2} y_{(u)}^T \left({\bf \bm{\rho}}^{-1}- \mbox{diag}(\bm{\rho} ^{-1})\right) y_{(u)} \right)\ .
\end{aligned}
\ea
Here $y_{(u)}=(\Phi_1^{-1}(u_1),\Phi_1^{-1}(u_2),\dots,\Phi_1^{-1}(u_n))^T$
and
\be
\mbox{diag}(\bm{\rho} ^{-1})= 
\begin{pmatrix} 
\begin{array}{cccc}
\sigma_1^{-2} & 0 & \cdots & 0 \\ 
0 & \sigma_2^{-2} & \cdots & 0 \\
\vdots & \vdots & \ddots & \vdots \\ 
0 & 0 & \cdots & \sigma_n^{-2} 
\end{array}
\end{pmatrix}
\ee
is the diagonal part of $\bm{\rho} ^{-1}$. One can further verify that, as we expect,  
\ba
f(\delta_1,\delta_2,\dots,\delta_n) 
=c_{\rm G}(u_1, u_2, \dots, u_n){\displaystyle\prod_{i=1}^{n} {f(\delta_i)}} \no \\
=\frac{1}{\sqrt{(2\pi)^n\det {\bf \bm{\rho}}}} \exp \left( -\frac{1}{2} \delta_{(u)}^T {\bf \bm{\rho}}^{-1} \delta_{(u)} \right)\ .
\label{jpdf}
\ea
Since the copula is invariant under local monotonically increasing transformation, 
\begin{equation}
\label{eq:defgc2}
C_{\rm G}(u_1, \dots, u_n) = \Phi_{\bm{\rho}',n} \left(\Phi^{-1}(u_1),
\dots, \Phi^{-1}(u_n)\right)\ .
\end{equation}
Now $\Phi_{\bm{\rho}',n}$ is the JCDF of $\delta_i/\sigma_i$ ($i=1, 2, \dots,n$) and   $\Phi$ is the one-point Gaussian CDF with unit variance.  The Gaussian copula density can then be simplified to 
\begin{equation}
\begin{aligned}
\label{defgcdnpt}
c_{\rm G}(u_1, u_2, \dots, u_n)=\frac{1}{\sqrt{\det {\bf \bm{\rho}'}}} \exp \left( -\frac{1}{2} y_{(u)}^T ({\bf \bm{\rho}'}^{-1}- \bm{I}) y_{(u)} \right)\ .
\end{aligned}
\end{equation}
Any Gaussian copula/copula density  can be written in form of Equation (\ref{eq:defgc2}) and Equation (\ref{defgcdnpt}). For simplicity, we use these two forms to illustrate Gaussian copula (densities) in this work.

\section{Testing the Gaussian Copula Hypothesis} 
\label{sec:def}
The Gaussian copula hypothesis (GCH, \citet{Scherrer2009}) states that, despite strong non-Gaussianity of one-point PDF $f(\delta)$,  the copula of the cosmological matter density field is Gaussian,  as  expressed by Equation (\ref{eq:defgc2}). It also predicts that the covariance matrix ${\bm \rho}$ is the one of $y=\Phi^{-1}(u=F(\delta))$. Namely, $y$ is the Gaussianization of $\delta$ such that its one-point PDF is Gaussian.  For the two-point copula, the only free parameter there is $r$, i.e. the cross correlation coefficient between $y_1$ and $y_2$.  The normalized covariance matrix
\be
\bm{\rho}=
\begin{pmatrix} 
\begin{array}{cc}
1 & r \\
r & 1 
\end{array}
\end{pmatrix}\ .
\ee

We test GCH with a $\Lambda$CDM N-body simulation. The simulation was run with $3072^3$ particles in a box of side length $600\  {\rm Mpc}/h$, and a
flat cosmology specified by $\Omega_m = 0.268$, $\Omega_\Lambda = 0.732$,
$H_0 = 71\ \rm{km}\ s^{-1}{Mpc}^{-1}$, $\sigma_8 = 0.83$, $n_s = 0.968$.  The details of the simulation are described in \cite{2007ApJ...657..664J} \& \cite{2019SCPMA..6219511J}.
The density fields are sampled at redshifts $z=17,5,1,0$ with the pixel size $1 {\rm Mpc}/h$. The mean number of particles per pixel is $\sim 134$, so we can safely neglect the effect of shot noise. As explained earlier, the copula may be misleading in revealing the non-Gaussianity. So we measure both the copula and the copula density.

We restrict our investigation on the two-point copula (densities), which can be measured from the JCDF and JPDF. 
To measure the joint distributions, we sample $n=600^3$ of $\delta_1$ at position ${\bf x}_1$ and the associated $\delta_2$ at position ${\bf x}_2={\bf x}_1+{\bf s}$. ${\bf s}$ is the  pair separation vector, and we investigate the cases of $s=2{\rm Mpc}/h$ and  $6{\rm Mpc}/h$, respectively. 
We follow the procedure described in \cite{Scherrer2009}, and utilize the transformation invariant property of copulas. We rank $\delta_1$ and adopt the monotonic transformation $y_1=R_1(\delta_1)/n$. Namely, the lowest $\delta_1$ is mapped to $y_1=1/n$ and the highest $\delta_1$ corresponds to $y_1=1$. We do the same for $\delta_2$ to obtain $y_2$. $y$ has the uniform PDF, $F(R_i(\delta_i)/n)=R_i(\delta_i)/n,\ f(R_i(\delta_i)/n)=1$. Then
\begin{equation}
 C_{\rm D}(R_1(\delta_1)/n, R_2(\delta_2)/n)=F(R_1(\delta_1)/n,R_2(\delta_2)/n)\ ,
\label{Cdjpdf_n2}
\end{equation}
\begin{equation}
 c_{\rm D}(R_1(\delta_1)/n, R_2(\delta_2)/n)=f(R_1(\delta_1)/n,R_2(\delta_2)/n)\ .
\label{cdjpdf_n2}
\end{equation}
Here, the subscript ``D'' denotes the data(simulation). 
In other words, the joint distributions of density ranks (divided by the number of points) give the 2-point copula (density), which is called the ``empirical copula".

\subsection{Direct Comparison}
Figure \ref{Cuv} shows the copula at $4$ redshifts. One finding is the lack of evolution in redshifts,  implying that the rank order of the density field roughly conserves under gravitational evolution \citep{1992MNRAS.254..315W}. We also over-plot the Gaussian copula predicted by GCH. The curves almost completely overlap with the simulation result, for all the $4$ redshifts and two spatial separations investigated. This  confirms the finding of \cite{Scherrer2009}. However, as we argue earlier, this is very misleading. To demonstrate this point, we over-plot the copula of vanishing $\langle \delta_1\delta_2\rangle$. From the viewpoint of LSS, this one is fundamentally different. Nevertheless, it overlaps with the simulated one when $u,v\rightarrow 0$, or $u,v\rightarrow 1$. Therefore, even tiny difference in the copula may lead to significant difference in LSS statistics.

Figure \ref{cuv} shows the copula density. In contrast to the case of copula, now departures from Gaussianities are clearly revealed (black solid curves versus black dash curves), at low redshifts or small separation. The next step is to quantify its impact on commonly used LSS statistics.

\subsection{GCH Induced Bias in LSS Statistics}
The two-point copula density determines all correlation functions of the following form,
\ba
\xi_{mn}=\langle\delta_1^m\delta_2^n\rangle&=&\int_{-1}^\infty \int_{-1}^\infty\delta_1^m\delta_2^n f(\delta_1,\delta_2)d\delta_1 d\delta_2\no \\
&=&\int_0^1 \int_0^1 \delta_1^m\delta_2^n c(u_1,u_2)du_1du_2\ .
\label{cdjpdf_n213}
\ea
The GCH fixes $c$ and therefore makes a unique prediction of $\langle\delta_1^m\delta_2^n\rangle$. Inaccuracies in GCH can then be quantified by the bias in $\langle\delta_1^m\delta_2^n\rangle$, with respect to the simulated (true) value.   As shown in Figure \ref{GCHbias1}, $\xi_{mn}$ predicted by GCH is accurate only at high redshift. Significant bias has developed even at $z=5$. Therefore despite (almost) invisible deviation from GCH in copula, the induced bias in $\xi_{mn}$ can be significant.

\subsection{The Alternative Gaussian Copula Approximation}

\begin{figure}[ht!]    
\centering
\includegraphics[width=9cm]{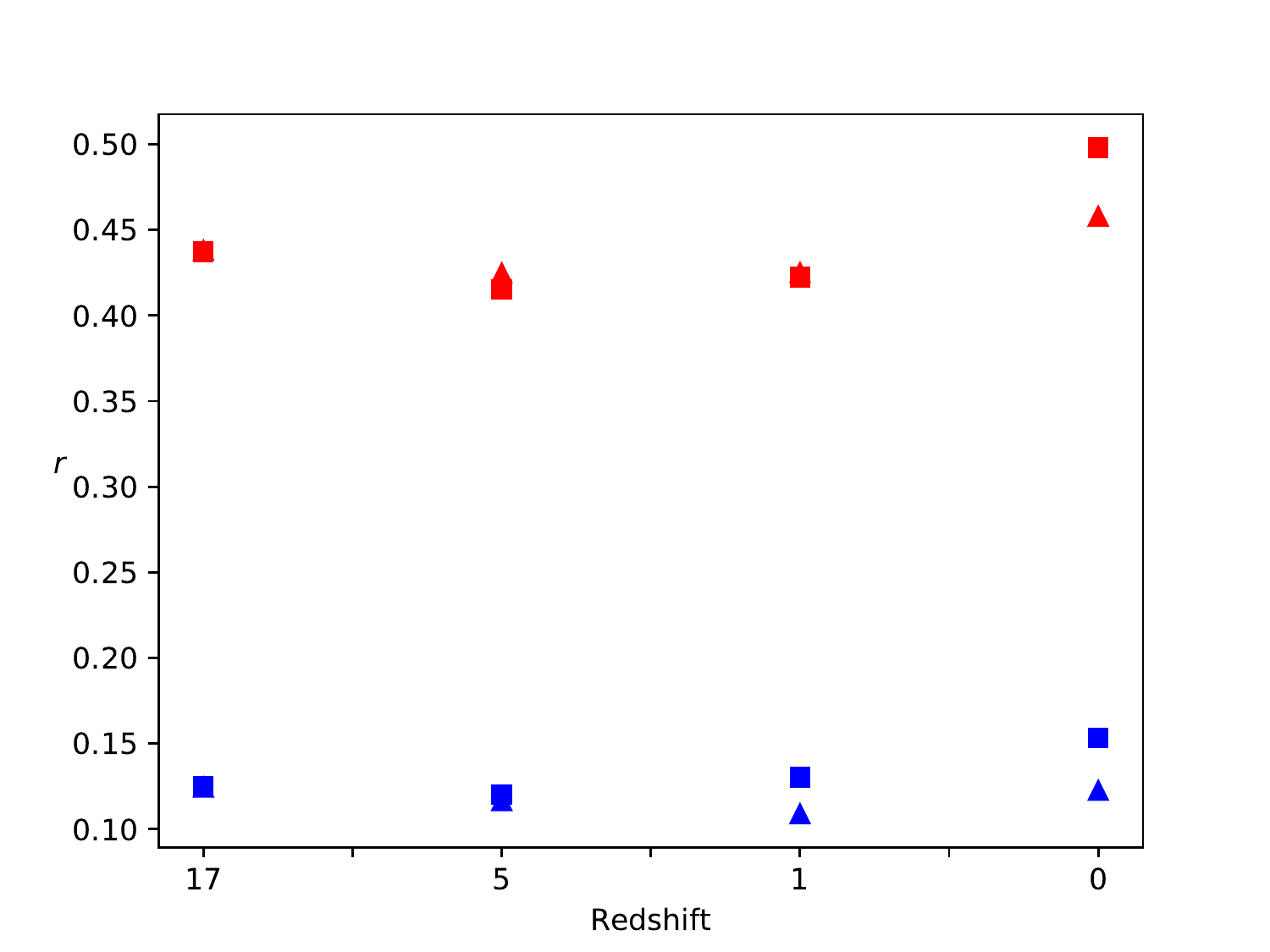}    
\caption{The results of $r$ of the Gaussian copula at redshift 0, 1, 5, 17 and at separation $2 \mathrm{Mpc}/h$ (the red points) and $6 \mathrm{Mpc}/h$ (the blue points). The triangular points are estimated from GCH, and the square points are estimated from the alternative Gaussian copula approximation.}
\label{Rvalues}
\end{figure} 

\begin{figure*}[ht!]
\centering
\includegraphics[width=19cm]{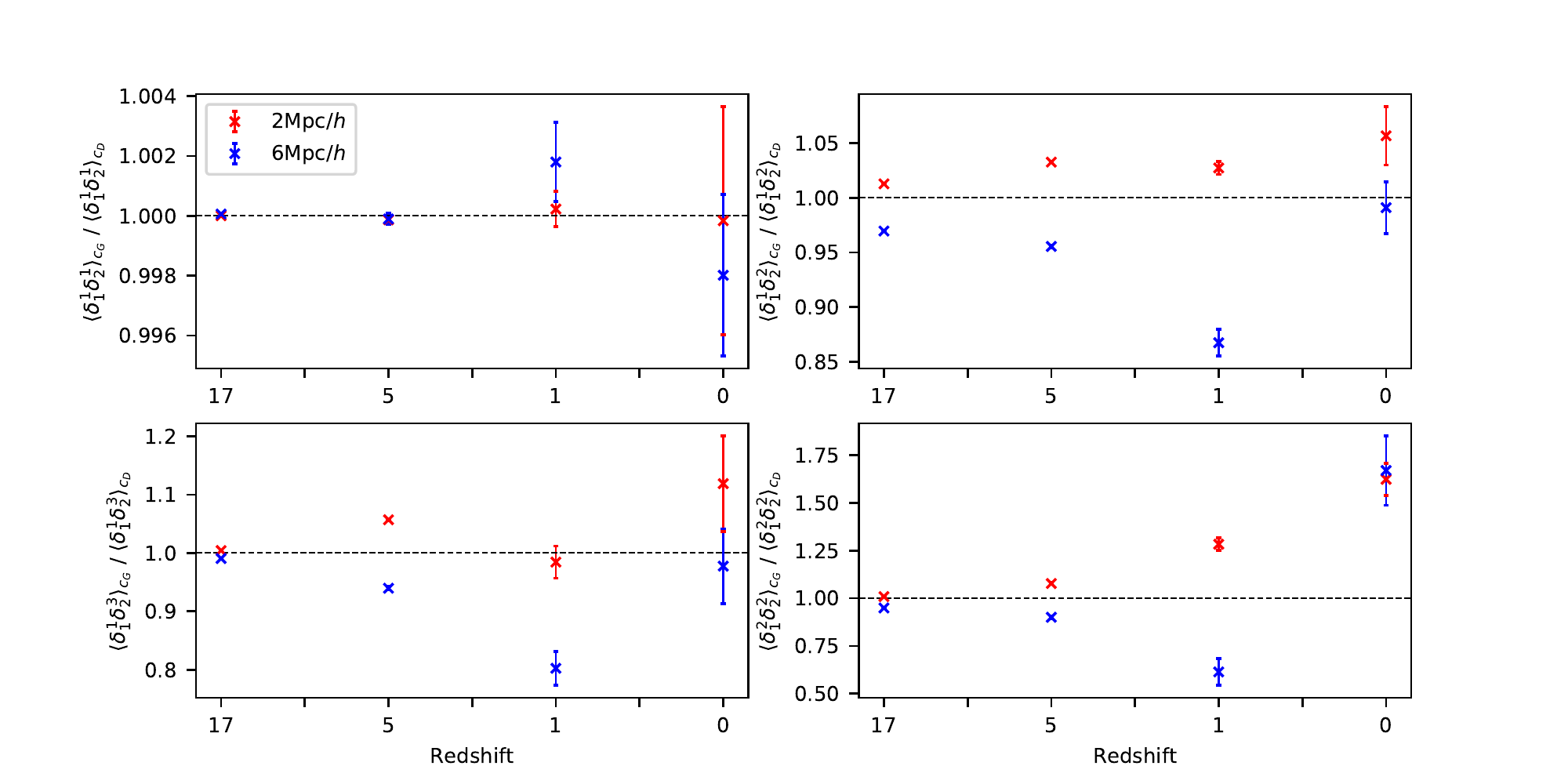}  
\caption{Same as Figure \ref{GCHbias1}, but for the Gaussian copulas determined by the alternative Gaussian copula approximation.}
\label{GCHbias2}
\end{figure*}

A surprising finding above is that GCH even fails to predict $\langle \delta_1\delta_2\rangle $ at low redshift. Since the prediction is completely fixed by $r$ in the covariance matrix ${\bm \rho}$, this motivates us to check whether we can choose another $r$ to improve the prediction of not only $\langle \delta_1\delta_2\rangle $, but $\langle \delta^m_1\delta^n_2\rangle $ in general.  $r$'s in GCH are fixed by the field $\Phi^{-1}(F(\delta))$\footnote{{\color{black}{{According to the transformation invariance of the copula, the GCH indicates that $r$ equals the correlation coefficient of $\Phi^{-1}(F(\delta_1)),\Phi^{-1}(F(\delta_2))$.We also checked that the value of $r$ determined by $\Phi^{-1}(F(\delta))$ is same to that if we follow the Spearman rank correlation procedure in \cite{Scherrer2009}}.}}}. We show them in Figure \ref{Rvalues}.  An alternative $r$ can be fixed by requiring that $\langle \delta_1\delta_2\rangle$ predicted by Equation (\ref{cdjpdf_n213}) agrees with the simulated one. Such $r$'s are also shown in Figure \ref{Rvalues}. The two sets of $r$ do show visible difference at low redshifts.  To distinguish from the GCH copula, we call the copula with the new set of $r$ as the ``alternative Gaussian copula approximation''.

Copula under the alternative Gaussian copula approximation gives unbiased result of $\langle\delta_1\delta_2\rangle$. However, they do not give better match for the copula densities (Figure \ref{cuv}). 
Furthermore, they improve the accuracy of predicted $\langle\delta_1^m\delta_2^n\rangle$ ($m+n>1$), but not significantly (Figure \ref{GCHbias2}).  Biases in $\langle\delta_1\delta_2^2\rangle$ vary from 5\% to 15\%.
Biases in $\langle\delta_1^m\delta_2^n\rangle$ ($m+n=4$) are larger, ranging from $10\%$ to  $\sim 60\%$. Such biases are too large for precision cosmology. Therefore even the alternative Gaussian copula approximation has limited usage in precision cosmology.


\section{Summary}
 We have revealed the otherwise hidden non-Gaussianity of the copula of the (3D) cosmological matter density field, via the copula density statistics and the accuracy in the predicted $n$-point correlation functions. The found non-Gaussianity shows that the nonlinear statistics of the 3D density field is more complicated than the Gaussian Copula hypothesis suggests. This further verifies our previous finding that the  $y$ field after local Gaussianization has detectable non-Gaussianity. One remaining question is the information encoded in the non-Gaussian part of the $y$ field, and another question is whether we can conveniently describe and capture such non-Gaussianity. These are for future works. On the other hand, Gaussianization of 2D density field (e.g. the weak lensing convergence field) is much more accurate, and has valuable applications \citep{2011MNRAS.418..145J,2014JCAP...04..004M,2011PhRvD..84b3523Y, 2012PhRvD..86b3515Y,2016PhRvD..94h3520Y,2020arXiv200110765C}.  
 
 {\color{black}{{The Copula is a promising tool  because of its advantageous mathematical properties, but its misuse can be misleading.
For example, the misuse of Gaussian copula in econometric modeling was blamed for the 2008 global financial crisis.
To make full and correct use of copulas, we need knowledge and judgment  beyond that used in traditional statistical measures. There are methodology developed by mathematicians for other applications that we can draw lessons from.
For example, in geology, \cite{GRALER201487}  found that the vine copula allow to include extremal behaviour of a spatial random field and to capture the distribution of heavily skewed spatial random field, where Gaussian copula failed.
 In structural engineering, \cite{WANG201875} found that, while the specification of random fields in terms of the marginal distributions and correlation structure is incomplete, the non-Gaussian dependence structure is a real phenomenon in engineering practice and they found the D-vine copula are more suitable for representing one-dimensional stationary random field.
In ecology, \cite{PRATES2015382} transform the margins of a Gaussian Markov random field to desired marginal distributions, which accommodate asymmetry and heavy tail needed in many ecological circumstances.}}}

\section{Acknowledgements}
This work was supported by the National Key Basic Research and Development Program of China (No. 2018YFA0404504), the National Science Foundation  of China (11621303, 11653003, 11773048, 11890691).

\bibliography{mybib}
\end{document}